\documentclass[
 reprint,
 amsmath,amssymb,
 aps,
]{revtex4-2}

\DeclareMathOperator{\sinc}{sinc}
\usepackage{graphicx}
\usepackage{dcolumn}
\usepackage{bm}

\usepackage{makecell}
\usepackage{multirow}

\usepackage{hhline}

\begin{document}


\title{Approaching optimal microwave-acoustic transduction on lithium niobate using SQUID arrays}

\author{A. Hugot}
 \email{abel.hugot@neel.cnrs.fr}
\author{Q. Greffe}
\author{G. Julie}
\author{E. Eyraud}
\author{F. Balestro}
\author{J. J. Viennot}
 \email{jeremie.viennot@neel.cnrs.fr}
\affiliation{Univ. Grenoble Alpes, CNRS, Grenoble INP, Institut Neel, 38000 Grenoble, France
}%

\date{\today}

\begin{abstract}
\textbf{Electronic devices exploiting acoustic vibrations are ubiquitous in classical and quantum technologies. Central to these devices is the transducer, which enables the exchange of signals between electrical and acoustic networks. Among the various transduction mechanisms, piezoelectricity remains the most widely used. However, conventional piezoelectric transducers are limited to either small efficiencies or narrow bandwidths and they typically operate at fixed frequency. These limitations restrict their utility in many applications. Here we propose and demonstrate a robust strategy to realize piezoelectric microwave-acoustic transduction close to the maximal efficiency-bandwidth product of lithium niobate. We use SQUID arrays to transform the large complex impedance of wide-band interdigital transducers into 50 $\Omega$ and demonstrate unprecedented efficiency$\times$bandwidth $\approx$ 440 MHz, with a maximum efficiency of 62\% at 5.7 GHz. Moreover, leveraging the flux dependence of SQUIDs, we realize transducers with \emph{in-situ} tunability across nearly an octave around 5.5 GHz. Our transducer can be readily connected to other superconducting quantum devices, with applications in microwave-to-optics conversion schemes, quantum-limited phonon detection, or acoustic spectroscopy in the 4-8 GHz band.}
\end{abstract}   

\maketitle


Intense research efforts have been invested in acoustic devices recently, notably for classical and quantum conversion between microwave and optical domains \cite{Brubaker2022, Blesin2023, Jiang2023, Meesala2024, Weaver2024, Zhao2024}, manipulation of solid-state spin qubits \cite{Whiteley2019, Maity2022, Dietz2023}, or fundamental tests of quantum mechanics \cite{Schrinski2023}. In these technologies, the transducer is the indispensable circuit component that allows two-way transformation between electrical and acoustic signals. For many of the envisioned quantum technologies, this transducer must be reciprocal, have a high efficiency-bandwidth product, and ideally offer tunability for practical reasons \cite{Zeuthen2020}. Among the physical mechanisms used for transduction, the piezoelectric effect inherently mediates a reciprocal and linear conversion of electrical and acoustic signals at the same frequency. Piezoelectric materials have been integrated with superconducting microwave circuits for about a decade \cite{O'Connell2010, Gustafsson2014} and this idea has led to the deployment of the circuit quantum electrodynamics techniques in the acoustic domain \cite{Clerk2020}.

The intrinsic challenges for piezoelectric transduction can be understood by looking at the electrical impedance of a transducer \cite{Morgan2007, Hashimoto2009}. In general, this impedance is strongly frequency-dependent and shows real and imaginary parts much larger than 50 Ohms. The direct consequence is that any incoming electronic signal hits a mismatched component and is mostly reflected back. Without additional engineering, this makes transducers very inefficient and incompatible with many quantum applications. Several strategies exist to mitigate this issue. If the acoustic network is restricted to resonant modes for instance, sufficiently high quality factors can compensate for the weak transducer coupling strength. On an arbitrary acoustic network, with strong piezoelectric materials, it is possible to match the transducer impedance to 50$\Omega$ by careful design \cite{Ekstrom2017,Dahmani2019}. However, this approach only works at specific fixed frequencies and over narrow bandwidths (order of 1-10 MHz), making such transducers impractical in more complex devices. A more suitable idea is to use impedance matching techniques \cite{Morgan2007, Pozar}. Basic microwave impedance matching has been investigated recently in the context of microwave-to-optics conversion, by coupling superconducting microwave resonators to piezoelectric transducers \cite{Jiang2023,Meesala2024,Weaver2024}. However, these works still rely on weakly coupled discrete modes. Therefore, these microwave-acoustics transducers can only operate over narrow bandwidths and correspondingly moderate efficiency$\times$bandwidth on the order of 1 MHz so far.

\begin{figure*}[ht]
\includegraphics{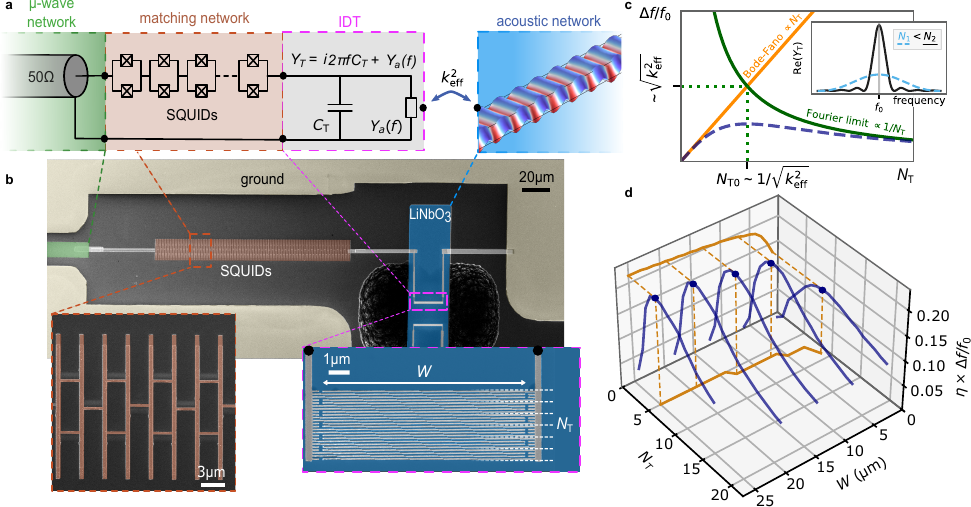}
\caption{\textbf{\label{fig1} The impedance-matched wide-band acoustic transducer.} \textbf{a-b}, Schematic and false color SEM micrograph of the wide-band bi-directional transducer. The interdigital transducer (IDT, grey) piezoelectrically excites Lamb waves in a suspended lithium niobate thin film (blue). The frequency-dependent admittance of the IDT is impedance-matched to the 50$\Omega$ microwave network (green) through an impedance transformer consisting of a chain of SQUIDs (brown).  \textbf{c} The fractional bandwidth of an impedance-matched transducer is constrained by the IDT's response (green, see inset) and by the Bode-Fano limit (yellow). The blue dashed curve illustrates the typical dependence of the fractional bandwidth around the optimal $N_T$. \textbf{d} The simulated efficiency times fractional bandwidth (\cite{SM}) of an optimally matched transducer as a function of $N_T$ for various values of the IDT's aperture $W$ (blue). The positions of the maxima are projected on the ($\eta \times \Delta f/ f_0$, $N_T$) and ($W$,$N_T$) planes (brown).
}
\end{figure*}

In this Article, we demonstrate transduction between 50 $\Omega$ microwave networks and acoustic networks that approach the intrinsic limits of the piezoelectric effect.  Our strategy does not rely on high-Q acoustic resonances. Instead, we use SQUID arrays as microwave impedance transformers. By exploiting the large piezoelectric coupling of lithium niobate thin films, we realize transducers that are both highly efficient and wide-band. With unidirectional IDTs, we demonstrate devices that reach 62 $\%$ efficiency and a corresponding efficiency$\times$bandwidth $\approx$ 440 MHz, more than two orders of magnitude larger than previously demonstrated \cite{Jiang2023,Meesala2024,Weaver2024}. In addition, the \emph{in-situ} flux tunability of our SQUID arrays can be exploited to make widely frequency-tunable transducers. When prioritizing tunability, we demonstrate transducers operating with efficiencies greater than 5 $\%$ in the 4-7 GHz band and a dynamical bandwidth of up to 740 MHz. When coupled to high-Q acoustic resonators, our transducers behave as tunable couplers, with high on/off ratios $\approx$ 50. The nonlinearity of our SQUID arrays is weak and enables the linear transduction of signal powers ranging from the quantum scale up to relatively large powers of -83 dBm. \\

\textbf{Targeting optimal transduction}
Interdigital transducers (IDTs) are the most common way to couple electrical signals with surface acoustic waves (SAWs) and Lamb waves. They are based on periodic arrangements of metallic electrodes on piezoelectric materials. As shown in Fig. \ref{fig1}(a), the electrical admittance $Y_T$ of an IDT is the parallel connection of a geometrical capacitance and a complex frequency-dependent acoustic part. Consequently, an IDT on its own cannot be matched to standard electrical networks over a large bandwidth (i.e. satisfy $1/Y_T=50\Omega$). Two general bounds can be identified when matching an IDT with an additional circuit. First, the frequency response of an IDT is given by the Fourier transform of its periodic arrangement of electrodes \cite{Morgan2007}, giving a center frequency $f_0$ and a bandwidth that scales with the inverse number of electrode periods $N_T$. Second, the Bode-Fano criterion \cite{Bode1945,Fano1950,Pozar} gives us a fundamental limit to how well we can match $Y_T$ using lossless matching networks. This criterion simplifies to a $\sim k_\textrm{eff}^2 N_T$ scaling, where $k_\textrm{eff}^2$ denotes the effective piezoelectric coupling coefficient of the IDT. This general reasoning, illustrated in Fig. \ref{fig1}(b), shows that there exists an optimum $N_{T0}$. We therefore get a general scaling law that determines the optimum $N_{T0}$ for externally matched IDTs. It also gives their maximum fractional bandwidth $\Delta f /f_0$ (at $N_{T0}$) which, remarkably, depends only on $k_\textrm{eff}^2$ \cite{SM}:

\begin{equation} \label{eq:N0}
N_{T0} \sim \frac{1}{\sqrt{k_\textrm{eff}^2}}
\end{equation}
\begin{equation} \label{eq:fracBW}
\frac{\Delta f}{f_0} \sim \sqrt{k_\textrm{eff}^2}
\end{equation}

\begin{figure*}[!ht]
\includegraphics{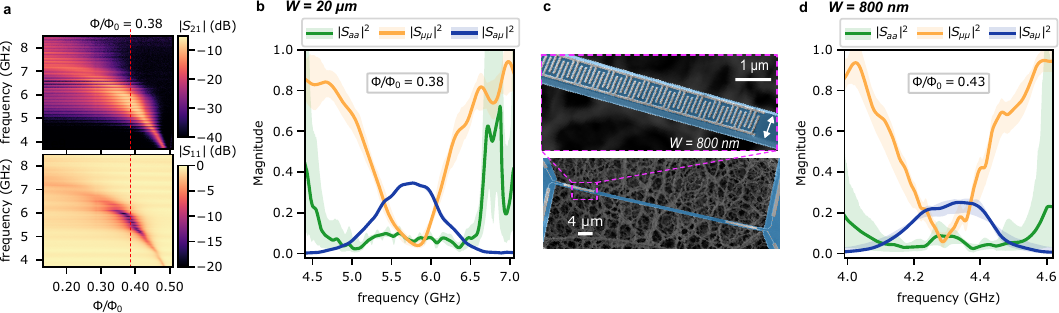}
\caption{\label{fig2} \textbf{Characterization of plane wave and quasi-single mode bi-directional transducers.} \textbf{a}, Measured transmission and reflection through a delay line consisting of two plane wave transducers as a function of frequency and magnetic flux threading the SQUIDs' loops.  \textbf{b}, Scattering parameters of the transducer after Fourier filtering the single transit signal from the raw data of \textbf{a}. Solid curves are mean values and shaded areas indicate 90\% confidence intervals (see \cite{SM}). \textbf{c}, False color SEM micrographs of a quasi-single mode IDT (grey) after release of the acoustic waveguide (blue).
\textbf{d} Same as \textbf{b} for the quasi-single mode transducer.}
\end{figure*}

The acoustic waveguides we employ here are suspended thin films of lithium niobate that support Lamb waves \cite{Arrangoiz2016}. We choose lithium niobate as a piezoelectric material because of its relatively high piezoelectric coupling and technological readiness. Based on this choice of waveguide and the known properties of lithium niobate, we obtain an estimate of $N_{T0}\approx 5$ and in turn, we estimate $Y_T$ which falls in the range $\Re(1/Y_T),\Im(1/Y_T) \approx 0.1 - 2$ k$\Omega$ \cite{SM}. High-impedance Josephson circuits are therefore particularly well suited to realize the matching network. We use impedance transformers made out of SQUID arrays, which are relatively simple and offer a large tunability. To verify the theoretical performance of a single transformer-IDT circuit, we perform finite element simulations to obtain a realistic model for $Y_T$. We then solve the transformer equations for $\eta = 1$ at the IDT's center frequency (see \cite{SM}) and numerically compute the fractional bandwidth of the circuit. The results shown in Fig. \ref{fig1}(d) indicate that for IDTs with a large aperture $W$ (i.e. IDTs that couple to plane waves in the 2D waveguides), there is an optimum point around the expected $N_T = N_{T0} = 5$, independently of $W$. For small $W$, $N_{T0}$ and $\Delta f/f_0$ begin to have a dependence on $W$ and deviate from equations (\ref{eq:N0},\ref{eq:fracBW}). This observation suggests that a more complex matching circuit might improve performances further. Here we opt for single SQUID impedance transformers regardless of the IDT aperture to retain simplicity and robustness.

\textbf{Experimental setup}
We experimentally demonstrate these results by fabricating acoustic Lamb waveguides on X-cut lithium niobate thin films. We choose an in-plane propagation angle rotated by $2.575$ rad from the Y crystallographic axis. In this orientation, the coupling of these IDTs with shear modes cancels out, and we mostly couple to the fundamental flexural ($A_0$) Lamb modes in our frequency band \cite{SM}. Similarly to previous works \cite{Arrangoiz2019}, we fabricate the superconducting microwave circuits away from lithium niobate (on high-resistivity Si) to minimize microwave losses, and we restrict the wiring on piezoelectric materials to the minimum outside the IDTs. We fabricate and characterize our SQUID arrays following the procedures of ref. \cite{Martinez2019, SM}. All the measurements presented here are performed at low temperatures (base temperature $\approx 30$ mK) within a  $50\Omega$-matched microwave network. We calibrate the microwave measurements \emph{in-situ}, with calibration standards measured at base temperature \cite{SM}. \\

\begin{figure*}[!ht]
\includegraphics{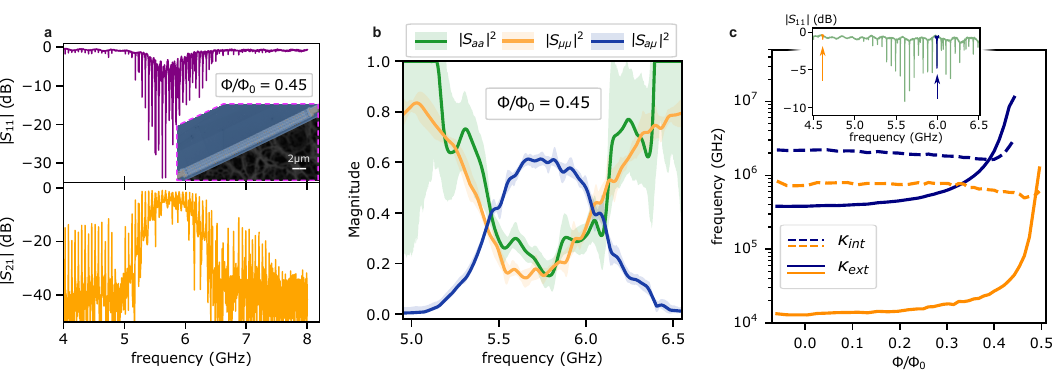}
\caption{\label{fig3} \textbf{Characterization of unidirectional transducers.} \textbf{a}, Reflection and transmission at fixed magnetic flux through a delay line consisting of two unidirectional transducers (UDT). The inset shows an SEM micrograph of such a transducer after release.  \textbf{b}, Scattering parameters of the UDT after Fourier filtering the single transit signal. Solid curves are mean values and shaded areas indicate 90\% confidence intervals (see \cite{SM}). \textbf{c}, Internal (dashed) and external (solid) loss rates as a function of applied magnetic flux for the two acoustic resonances highlighted in the inset (at $\phi = 0$).} 
\end{figure*}

\textbf{Bi-directional transducers}
We begin with devices consisting of two microwave ports, each connected to a transformer-IDT circuit and arranged in a delay-line geometry (Fig. \ref{fig1}.b shows half of this arrangement, see \cite{SM} for full devices). We characterize the transduction of acoustic plane waves as well as waves in nearly single-mode waveguides \cite{Dahmani2019}, with apertures $W$ of 20 µm and 800 nm respectively. These delay lines are not terminated acoustically and we use split electrodes \cite{Morgan2007}, so that waves traveling through an IDT leave the waveguide with negligible coherent backscattering. Consequently, there are no resonance conditions along the propagation direction. From microwave scattering measurements, we extract the transduction efficiency $\lvert S_{a \mu} \rvert ^2$, the microwave $\lvert S_{\mu \mu} \rvert ^2$ and acoustic mismatch $\lvert S_{a a} \rvert ^2$, shown in Fig. \ref{fig2}(c,d) \cite{SM}. For plane-wave transducers, we find a bandwidth of 900 MHz around a center frequency of 5.8 GHz. Thus the fractional bandwidth, $\Delta f/f_0 \approx 0.16$ approaches the value of 0.21 extracted from the simulations of Fig. \ref{fig1}(d). For single-mode waveguides, the measured bandwidth also reaches a large value of 270 MHz around 4.34 GHz corresponding to a fractional bandwidth, $\Delta f/f_0\approx 0.06$. While we expect a diminution from Fig. \ref{fig1}(d), this result is below the simulated value of $0.12$.

Fig. \ref{fig2}(c) shows that the return loss (i.e. mismatch) on the acoustic port $\lvert S_{aa} \rvert^2$ increases to $\approx$ 0.1 at the center frequency of the transducer, around 5.8 GHz. This is due to the three-port character of the IDTs used here: they emit acoustic waves that propagate in both opposite directions therefore coupling two acoustic ports to one microwave port. At $\approx$ 5.8 GHz, the microwave port is well matched ($\lvert S_{\mu \mu} \rvert^2 < 0.1$) and creates a load on the acoustic ports that raises acoustic reflections (theoretically, $\lvert S_{aa} \rvert^2 = 0.25$ for a perfect symmetric three-port device). Likewise, the transduction efficiencies $\lvert S_{a \mu} \rvert^2$ shown in Fig. \ref{fig2} correspond to the scattering between the microwave port and only one of the acoustic ports. $\lvert S_{a \mu} \rvert^2$ is therefore by definition limited to a maximum of $50\%$ for bi-directional IDTs. \\

\textbf{Unidirectional transducers}
In order to push for higher efficiencies, we design unidirectional IDTs for plane waves, by placing acoustic mirrors just behind bi-directional IDTs (UDTs). Wide-band acoustic mirrors can be realized for Lamb waves by making free-ended waveguides \cite{Sarabalis2020}. This contrasts with SAWs, for which unidirectional IDTs exist but are limited to narrow band properties \cite{Ekstrom2017,Qiao2023}. The waves emitted by such a transducer result from the constructive interference between the waves emitted in one direction, and those emitted in the opposite direction and reflected by the mirror. The interference condition depends on the distance to the free boundary and we optimize this distance using finite element simulations as well as measurements \cite{SM}. These interferences reduce the bandwidth of the IDT by a factor close to two but increase the piezoelectric terms of the admittance $Y_T$ by the same factor. Overall, the admittance of such unidirectional IDTs resembles that of a standard IDT but with an effective $N_T^\textrm{eff}\approx 2 N_T$. To maintain $N_T^\textrm{eff}\approx 5$, we further reduce the number of physical fingers, as shown in Fig. \ref{fig3}(a). In the continuous wave transmission and reflection measurements of the delay line shown in Fig. \ref{fig3}(a), there are now several high-Q resonances. These resonances correspond to the harmonic (Fabry-Perot) modes of the free-ended acoustic waveguide. By selecting the single acoustic transit signal in the time domain  \cite{SM}, we filter out the acoustic resonances to obtain the scattering parameters of the circuit shown in Fig. \ref{fig3}(b).
 We find that, as expected, the maximum transduction efficiency, $\lvert S_{a \mu} \rvert^2 = $ 62\%, is improved by a factor close to 2 compared to the bi-directional IDT, thus exceeding the 50\% threshold needed for the linear transduction of quantum states \cite{Gyongyosi2018}. The fractional bandwidth of the UDT is still large, $\Delta f/f_0\approx 0.13$ (711 MHz around 5.69 GHz), and this transducer achieves an unprecedented efficiency$\times$bandwidth $\approx$ 440 MHz.
Our SQUID transformer-IDT circuit remains linear until the Josephson non-linearity starts to play a role \cite{Planat2019} and we find that the efficiency deviates by 1dB at around -83 dBm at the input of the SQUID array.

The built-in flux tunability of the SQUID transformer allows us to use our devices as tunable couplers for acoustic modes. To demonstrate this functionality, we use devices similar to that shown in Fig. \ref{fig3}(a), but with a single transducer.
 We now keep all the acoustic transits to analyze the Fabry-Perot resonances shown in the inset of Fig. \ref{fig3}(c). Changing the flux $\phi$ in the SQUID loop, we tune the external coupling $\kappa_e$ of the acoustic modes to the microwave network. We control $\kappa_e$ over 1.5 to 2 decades and up to large values in excess of 10 MHz for modes close to the IDT center frequency. At higher fluxes, the acoustic modes overlap and extraction of the couplings becomes unreliable. We speculate that the slight increase in internal acoustic losses $\kappa_i$ around $\phi/\phi_0 = 0.44 $ is due to the significant participation of the SQUID transformer to the normal mode of the system in this regime. (see loss budget below).\\

\begin{figure}[ht]
\includegraphics{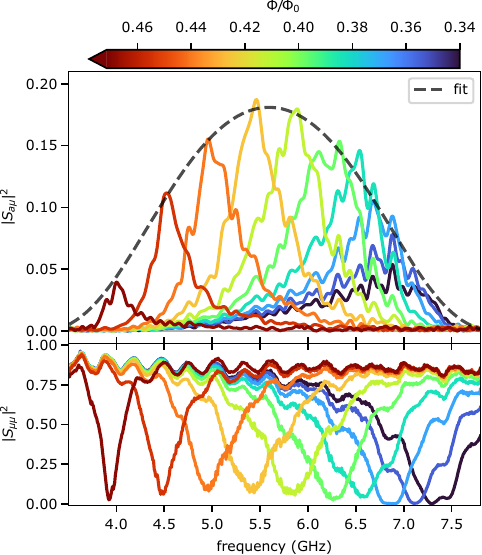}
\caption{\label{fig4} \textbf{In-situ tunability in the 4 - 7 GHz band.} Transmission and reflection coefficients of an IDT optimized for tunability for varying magnetic flux. The envelope of the transmission coefficient is fitted using an electrical model taking into account losses as a parasitic conductance $G_{\text{loss}}$ (see \cite{SM}). }
\end{figure}

\textbf{Widely tunable transducers}
In addition to being usable for the linear transduction of quantum states, we envision that our device could be advantageously exploited for sensitive acoustic spectroscopy. For such experiments, operation over a wide frequency band is essential.
 To achieve this goal, we design transducers with the largest possible frequency tunability. Therefore, we leave the optimal IDT design ($N_{T0}\approx 5$) and fabricate regular bi-directional IDTs with fewer fingers ($N_T=3$) to push away the Fourier limit. 
 
The scattering parameters of such a transducer, shown in Fig. \ref{fig4}, demonstrate a tuning range that nearly covers the C band (4-8 GHz). As shown by the substantial dips in the reflection $\lvert S_{\mu \mu} \rvert^2$, the device can be closely matched to 50 $\Omega$ over the band. Close to the IDT center frequency $f_0 = 5.6$ GHz, these dips in reflection correspond to large peaks in the efficiency $\lvert S_{a \mu} \rvert^2$. However, these peaks are increasingly reduced and shifted in frequency as we move away from the $f_0$. This can be explained by taking into account parasitic loss channels with the addition of a conductance $G_{\text{loss}}$ in parallel of the IDT. As the operation frequency is tuned away from $f_0$, the participation of the IDT conductance to the total conductance decreases and leads to reduced efficiency. We fit the envelope of the flux dependent $\lvert S_{a \mu} \rvert^2$  (dashed line) to extract $1/G_{\text{loss}}  = 8.4 \text{ k}\Omega$ \cite{SM}. Overall, we achieve a transduction efficiency exceeding 5 \% in the 4.1 to 7.1 GHz band. This frequency band is particularly interesting because it matches that of most superconducting quantum circuits.\\

\textbf{Loss budget}
While computing $G_{\text{loss}}$ can give us an indication on the amount of losses in our transducers, it does not explain the source of these inefficiencies. We report the analysis of these loss channels for the case of the unidirectional transducer presented in Fig. \ref{fig3}(a,b). The analysis for the other types of devices is given in \cite{SM}. The power impinging on the transducer can either be reflected (16 \%), transmitted as acoustic waves (61\%), or lost (23\%).  From the loss rate of the acoustic resonances, we can estimate acoustic propagation losses to contribute 3.2\% at $P_{\text{in}} = -108$ dBm (same power as in Fig. \ref{fig3}(a,b)). Beam-steering is estimated to contribute to 1.8 \%. The internal efficiency of our transducer can therefore be quoted as 66 \%. We note that acoustic losses increase up to 12 $\%$ at $P_{\text{in}} = -123$ dBm. In independent measurements on bare SQUID chains, we find internal quality factors of about 300, typical for such circuits \cite{Planat2019}. From this measurement, we estimate a contribution to the total amount of losses below 1 \%. We saw no increase in losses down to an input power of -123 dBm corresponding to $\approx$ 0.1 photons in the SQUID array. Finally, by analyzing the electromagnetic crosstalk through the delay line, we can estimate EM radiation to contribute to at least 3.8\% to the inefficiency. This leaves about 15\% of uncharacterized losses. We speculate that a significant part of these losses arise from the electrical wiring running over the lithium niobate, causing parasitic acoustic emission. One way to reduce these losses is to further reduce the length of this wiring, following the strategies recently used in other piezoelectric transducers \cite{Jiang2023,Meesala2024,Weaver2024}. \\

\textbf{Conclusions and outlook}
We have demonstrated a general method for designing optimal piezoelectric transducers based on IDTs and SQUID impedance transformers. Such transducers could be integrated within microwave-to-optics converters that rely on acoustic mediation. When designed rather for a large flux tunability, transformer-IDT circuits could be used to perform the acoustic spectroscopy of quantum systems that are not well coupled to electromagnetic fields. Examples of such systems are spins and high-frequency acoustic modes in two-dimensional materials. Thanks to their 50$\Omega$-matched microwave impedance, our circuits are directly compatible with existing Josephson traveling wave amplifier \cite{Ranadive2022} and single microwave photon detectors \cite{Wang2023}. Using these tools would allow for nearly quantum-limited linear detection of acoustic fields as well as phonon counting. \\

\textbf{Acknowledgements}
We acknowledge precious inputs from N. Roch and L. Planat on the nanofabrication technique of bridge-free Josephson junction arrays. The samples were fabricated in the clean room facility of Néel Institute (Grenoble) and we thank all the clean room staff for their assistance. We acknowledge L. Del-Rey, J. Jarreau, and D. Dufeu for their help in the installation and maintenance of the cryogenic setup. We thank A. Reinhardt, M. Tomasian and the members of the superconducting quantum circuits group at Neel for helpful discussions. This work was supported by the French National Research Agency (ANR) through the project MagMech (ANR-20-CE47-0004-01). Q. G. acknowledges financial support by the ANR agency under the France 2030 plan, with reference ANR-22-PETQ-0003. \\

\textbf{Author contributions}
A.H. and J.J.V. designed the experiment. A.H. fabricated the devices with inputs and support from G.J.. A.H. realized the measurements and performed the data analysis with the help of Q.G. and J.J.V.. E.E. provided support with the cryogenics of the experiment. F.B. and J.J.V. supervised the project. A.H. and J.J.V. wrote the manuscript with inputs from all authors. \\

\textbf{Data availability}
The data supporting the findings of this paper is available from the corresponding authors upon request.

\bibliography{BiblioHugot2024}

\clearpage

\begin{widetext}
 \centering
 \large \bfseries{Supplementary Materials for: Approaching optimal microwave-acoustic transduction on lithium niobate using SQUID arrays}\\[1.5em]
\end{widetext}

\tableofcontents

\section{Derivation of the scaling of the fractional bandwidth for externally matched IDTs}
The electrical admittance $Y_T$ of interdigital transducers (IDTs) on piezoelectric thin films are well approximated by the following model, widely used for surface acoustic wave IDTs \cite{Morgan2007}:
\begin{align} 
    Y_T(f)  &= G_a(f) + i B_a(f) + i 2 \pi f C_T \notag \\
     G_a(f) &= G_0  W  N_\text{T}^2 f_0 \sinc ^2 \left(\pi N_\text{T} \frac{f-f_0}{f_0} \right) \label{eq:YT}\\ 
    B_a(f) &= G_0 W  N_\text{T}^2 f_0 \frac{\sin \left(\pi N_\text{T} \frac{f-f_0}{f_0} \right)  -  \pi N_\text{T} \frac{f-f_0}{f_0} }{2\left(\pi N_\text{T} \frac{f-f_0}{f_0} \right)^2} \notag
\end{align}
with:
$C_T$ is the IDT capacitance, $W$ its acoustic aperture, $N_\text{T}$ the number of electrode pairs, $f_0$ its center frequency. We define: $\{ G_0 = 8 k_\text{eff}^2 C_\textrm{T0}    , \text{ } C_\textrm{T0} = C_T/( W N_\text{T}) \}$ where $k_\text{eff}^2$ is the piezoelectric coupling strength.  These parameters $G_0$ and $C_\textrm{T0}$ are, in principle, independent of $W$ and $N_\text{T}$ and are thus practical to analyze and design IDTs with various geometries.

From an electrical perspective, the power dissipated in $G_\textbf{a}$ corresponds to the power emitted as acoustic waves while the imaginary part $B_\textbf{a}$ ensures the causality of the IDT impulse response.




The $\sinc$ shape of $G_\textbf{a}$ is directly related to the the Fourier transform of the charge distribution in the IDT's electrodes \cite{Morgan2007}. The spatial distribution of the electrodes will therefore characterize the IDT's frequency response. The pitch between electrodes will determine the IDT's center frequency $f_0$ (and the wavelength $\lambda$ of the acoustic waves to which it couples). In contrast, the transducer length (i.e. the number of electrode pairs $N_\text{T}$) will determine the width of the $\sinc$. The first limit on the fractional bandwidth of an impedance-matched IDT therefore comes from the IDT's geometry itself. The transducer's response cannot be larger than the width of the central lobe of $G_\text{a}$. From the position of the roots of the $\sinc$ function we can therefore derive:
\begin{equation}\label{eq:Fourier_lim}
    \frac{\Delta f }{f_0}^{\text{geom}} \propto \frac{1}{N_\text{T}} 
\end{equation}
This yields the green curve of figure 1b in the main text.

The second limitation stems from the impedance matching circuit. Harder to match loads ($Z_\text{load}$ and $Z_0$ are far apart) lead to smaller possible bandwidths. This can be formalized by the Bode-Fano criterion which states, that for a parallel RC circuit, the bandwidth of an arbitrary impedance matching circuit will be limited as follows \cite{Pozar}:
\begin{equation}\label{eq:Bode-Fano}
    \Delta f \ln \frac{1}{\Gamma} \leq \frac{1}{2 R C}
\end{equation}
where $\Gamma$ is the maximum reflection that we allow on the frequency band $\Delta f$. 
We can apply this limit to an IDT at its center frequency to get the yellow curve of Fig.~\textbf{1}b:
\begin{equation}
    \Delta f^\text{Bode-Fano} \propto \frac{G_{0} W N_\text{T}^2 f_0 }{C_\textrm{T0} W N_\text{T}} \propto k_\text{eff}^2 N_\text{T}
\end{equation}


Equating these two bounds on $\Delta f$ gives the position of the optimum $N_\text{T0}$ and yields equations (1) and (2).

\section{Computation of target designs and theoretical performances}
\subsection{Typical admittance of IDTs on lithium niobate thin films}
We perform COMSOL finite element simulations to obtain a more accurate admittance $Y_T$ for our IDTs. The IDTs are modeled in a quasi 2D (with a tiny aperture W and periodic lateral boundary conditions) and without feeding wires. We introduce a rescaling factor of 0.85 on the piezoelectric tensor, similarly to previous works \cite{Dahmani2019}. We fit the simulated admittance to the model above (equation \ref{eq:YT}) to obtain the parameters $\{ G_0, \text{ } C_\textrm{T0} \}$. A typical value for our devices is $G_0 \approx 1.6 \mathrm{e}{-10} $ S/m/$N_\text{T}$. In terms of impedance, we find $(1/Y_T) \approx 76 - i 408 \text{ }\Omega$  for a plane wave IDT with $W = 20$ µm and $N_\text{T} = 5$. The simulated impedances of all the IDTs presented in this paper are summarized in Supplementary Table \ref{SMtab:Zidt}.


\begin{table}[t]
    \setlength{\tabcolsep}{6pt} 
    \renewcommand{\arraystretch}{1.3} 
    \centering
    \begin{tabular}{|c|c|c|c|}
        \hline
        & $\Re(1/Y_T)$ & $\Im(1/Y_T)$ & $\lvert 1/Y_T \rvert$ \\
        \hline 
        plane-wave IDT & $76 \text{ }\Omega$ & $-408 \text{ }\Omega$ & $415 \text{ }\Omega$ \\
        
        UDT & $150 \text{ }\Omega$ & $-597 \text{ }\Omega$ & $616 \text{ }\Omega$\\
         
        single mode & $420 \text{ }\Omega$ & $-2572\text{ }\Omega$ & $ 2606\text{ }\Omega$\\
        
        widely tunable IDT& $106 \text{ }\Omega$ & $-1418\text{ }\Omega$ & $1421\text{ }\Omega$  \\
        \hline
    \end{tabular}\caption{\label{SMtab:Zidt} \textbf{Summary of the simulated impedance of all measured devices.}}
\end{table}

\subsection{Impedance transformer for complex loads} \label{section:transformer}
When a transmission line of length $l$ is connected to an arbitrary complex load $Z_\text{L} = R_\text{L} + i X_\text{L}$ it effectively behaves as an impedance transformer. The impedance seen at its input is given by \cite{Pozar}:
\begin{equation}\label{eq:transfo}
    Z_{in} = Z_\text{T} \frac{Z_\text{L} + i Z_\text{T} \tan \beta l}{Z_\text{T} + i Z_\text{L} \tan \beta l}
\end{equation}
where $Z_\text{T}$ and $\beta l$ are the transmission line's characteristic impedance and electrical length respectively. One can then choose $Z_\text{T}$ and $\beta l$ such as to impedance match the load to the $Z_0 = 50 \Omega$ input. Solving $\Gamma = \frac{Z_\text{in}-Z_0}{Z_\text{in}+Z_0} = 0$ for  $Z_\text{T}$ and $\beta l$, we get:
\begin{widetext}
\begin{align} \label{eq:sol_transfo}
    \begin{split}
        &Z_T  = \sqrt{\frac{-Z_0^2 R_\text{L} +Z_0  R_\text{L}^2 +Z_0  X_\text{L}^2 }{ R_\text{L} - Z_0}} \\
        &\beta l  = \arctan{ \left( - \frac{ R_\text{L}^2 + X_\text{L}^2 -  Z_\text{T}^2 + \sqrt{4  X_\text{L}^2 Z_\text{T}^2 + \left( Z_\text{T}^2 - R_\text{L}^2 - X_\text{L}^2 \right)^2 }}{2 X_\text{L} Z_\text{T} } \right)} 
    \end{split}
\end{align}
\end{widetext}

Knowing that $Z_\text{T} = \sqrt{\frac{L}{C}}$ and $\beta = \frac{1}{\sqrt{L C}}$, we then compute the required inductance $L$, capacitance $C$ and physical length $l$ of the transmission line.
The Josephson inductance of the SQUIDs is designed and verified by DC measurement of their normal state resistance through the Ambegoakar-Baratoff relation. The capacitance is fixed geometrically using EM simulations. We have verified that this design procedure is accurate by measuring the dispersion relation of long (several mms) arrays of SQUIDs \cite{Planat2019}. 

\begin{figure}[ht]
    \includegraphics{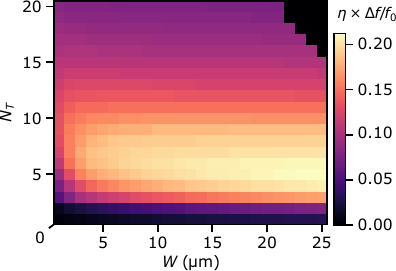}
    \caption{\label{figS1} \textbf{Simulation of the efficiency $\times$ fractional bandwidth} as a function IDT aperture $W$ and number of electrode pairs $N_\text{T}$ for a plane wave bi-directional IDT.}
\end{figure}
\subsection{Expected fractional bandwidth for various acoustic apertures}

In order to find the optimal transducer design, we compute the expected fractional bandwidth as a function of the IDT's acoustic aperture $W$ and number of electrode pairs $N_\text{T}$. 
For each $\{W, N_\text{T}\}$, we target a perfect match at the IDT center frequency where $Y_\text{T} = G_\text{a}(f_0)+i 2 \pi f_0 C_\text{T}$, and compute the transformer electrical length and characteristic impedance according to equations \ref{eq:sol_transfo}. Using these values as well as the complete frequency dependent IDT admittance in equation \ref{eq:transfo}, we compute $\lvert S_{11}\rvert^2$ and $\lvert S_{21}\rvert^2 = 1 - \lvert S_{11}\rvert^2$. Finally, we estimate the fractional bandwidth of the transducer by computing the full width at half maximum of $\lvert S_{21}\rvert^2$. For a perfectly matched IDT, the theoretical efficiency at $f_0$ should be 1 and we plot the colormap $\eta \times \Delta f / f_0$ of Supplementary Fig.~\ref{figS1}. In the main text, Fig.~1c is then plotted from this dataset.

As shown earlier in equations (1) and (2), the optimal $N_\text{T}$ and bandwidth (BW) should be independent of $W$. This holds true for large values of $W$, but deviations occur at smaller $W$, where the optimal point shifts toward higher $N_\text{T}$ and lower bandwidth. In this regime, the shift indicates that the Fourier limit becomes the dominant constraint, moving further away from the Bode-Fano limit. This suggests that the matching network becomes slightly less optimal as $W$ decreases.
\begin{figure*}[t]
\includegraphics{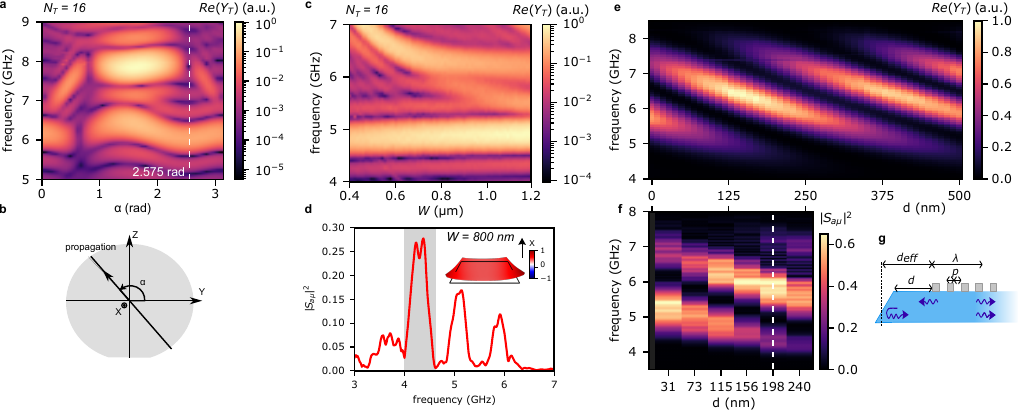}
\caption{\label{figS2} \textbf{Lithium niobate waveguides properties.} \textbf{a.} COMSOL simulation of $\Re (Y_T)$ for $N_\text{T} = 16$ as a function of $\alpha$, where $\alpha$ is the angle between the propagation direction of waves excited by the IDT and the LiNbO$_3$ crystal axes as defined in panel \textbf{b}. \textbf{c} COMSOL simulation of $\Re (Y_T)$ of an IDT with $N_\text{T} = 16$ on a single-mode waveguide as a function of the waveguide width $W$. \textbf{d} Measured conversion efficiency $\lvert S_{a \mu} \rvert ^2$ of an IDT on top of a $W = 800$ nm wide waveguide. The inset shows the simulated A0-like shape of the lowest frequency mode. \textbf{e.} COMSOL simulation of $\Re (Y_T)$ for an UDT as a function of the distance $d$ to the free edge mirror (as defined in panel \textbf{g.}). \textbf{f.} Measured conversion efficiency $\lvert S_{a \mu} \rvert ^2$ of UDTs with varying $d$. The dashed line indicates the devices that are shown in the main text.}
\end{figure*}

\section{Lithium niobate waveguides properties}
\subsection{Plane wave IDTs}

As a 3-port device can never be matched on all ports at once, it is necessary to ensure that an IDT does not couple to more than one acoustic mode. Narrow-band IDTs usually have responses to different propagating modes that are spectrally well separated. However, for wide-band applications, there may be instances where the IDT's response to distinct modes (such as A0 and SH0) overlaps, leading to significant transduction into undesired modes. In order to avoid such problems, we exploit the anisotropic properties of X-cut LiNbO$_3$. COMSOL simulations (Supplementary Fig.~\ref{figS2}a) show that depending on the in-plane propagation angle $\alpha$, our IDTs couple to two main acoustic modes in the 4-9 GHz range: a low-frequency flexural A0 mode and a high-frequency SH0 mode. For an angle $\alpha = 2.575$ rad with the Y crystallographic axis (see Supplementary Fig.~\ref{figS2}b), the coupling to SH0 modes vanishes and the IDT couples to the A0 mode only. Here the simulation was performed for a narrow band IDT with $N_\text{T} = 16$ to properly resolve the different modes. This decoupling from SH0 modes allows for the design of wide-band IDTs at the optimal $N_\text{T0}$ that can be matched both on the microwave and acoustic ports. All devices presented here are therefore realized at this angle.

\subsection{Single-mode waveguides and IDTs}

In addition to plane wave devices, we fabricate IDTs transducing into acoustic waveguides where waves are confined in all transverse directions. In this case, by reducing the width of the waveguide to a size comparable to the acoustic wavelength $\lambda$, it is possible to restrict acoustic propagation to a single mode. 

We simulate such a waveguide in COMSOL while varying its width $W$ and plot the IDT's response in Supplementary Fig.~\ref{figS2}c. 
Multiple modes are resolved: a low-frequency A0-like mode whose frequency remains largely independent of $W$ (see inset of Supplementary Fig.~\ref{figS2}d), and two higher-order modes which are increasingly well separated from the A0 mode for smaller $W$.
For $W < 1$ µm, the separation between these modes should be sufficient so as to distinguish them.
To balance good frequency separation of propagating modes, avoid parasitic internal modes, and maintain feasible nanofabrication, we choose to realize acoustic waveguides with $W = 800$ nm.
Note that our fabrication process yields sloped sidewalls, as seen in the inset of Supplementary Fig.~\ref{figS2}d and Fig.~2c of the main text.
 Supplementary Fig.~\ref{figS2}d shows the envelope of the measured transduction efficiency $\lvert S_{a \mu} \rvert ^2$ of a single mode transducer over the frequency band of interest. Although their position and separation are not exactly the same as in the simulation, we distinguish three main peaks corresponding to the different modes. Below 4 GHz, we observe the presence of parasitic modes that may correspond to the internal IDT modes. The grey-shaded area highlights the frequency band plotted in Fig.~2d of the main text.

\subsection{Unidirectional IDTs}
In order to recover the backward-emitted wave, an acoustic mirror can be placed behind the IDTs. For Lamb waves, this can be achieved by etching a free edge immediately behind the IDT. A critical parameter is then the distance $d$ between the IDT and the mirror, as this will determine the dephasing between the forward emitted and reflected waves and thus influence whether their interference is constructive or destructive. To optimize the device performance in terms of efficiency and bandwidth, it is necessary to align these interferences with the IDT's response.

We perform COMSOL simulations to study the effect of varying the distance between the IDT and free edge (Supplementary Fig.~\ref{figS2}e). For an edge with a straight vertical sidewall, constructive interference at the center frequency $f_0$ occurs when the IDT is positioned half a pitch away ($d = p/2 = \lambda/16$).  In our case, the sloped sidewall effectively behaves as a straight sidewall positioned further back, as shown in Supplementary Fig.~\ref{figS2}g. However, the slope makes it impossible to place the IDT at the optimal position directly adjacent to the effective edge.  Instead, we target the second-best position, located at $\lambda/2$ from the effective edge. At this position, the acoustic admittance of a UDT is approximately equivalent to that of an IDT with $N_\text{T} \times 2$. 

In practice, accurately determining the location of the effective edge is challenging due to imperfections inherent in the fabrication process. 
We therefore fabricate a series of devices with varying distances to the free edge, sweeping the distance across six values spanning a range of $\lambda/2$. This ensures that at least one device is positioned within $\lambda/12$ of the ideal position. We plot the transduction efficiency of these devices in Supplementary Fig.~\ref{figS2}f and observe that the general behavior of UDTs agrees well with the simulations. The device presenting the best performance is located at $d=198$ nm of the edge and is the one shown in Fig.~3 of the main text.

\section{Fabrication}
\begin{figure*}[ht]
    \includegraphics{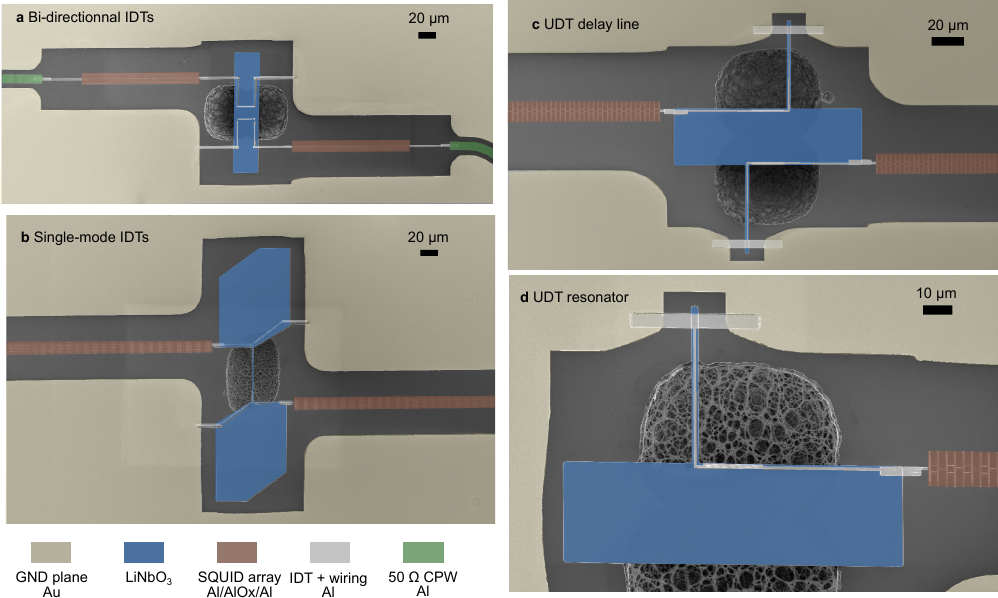}
    \caption{\label{figS3} \textbf{False color SEM micrographs of the complete devices.} \textbf{a.} Two bi-directional IDTs arranged in a delay-line geometry, zoomed-out view of Fig.\textbf{1}b of the main text. Same for single mode IDTs (\textbf{b.}) and UDTs (\textbf{c.}). A single UDT is used to couple to an acoustic resonator (\textbf{d.}). }
\end{figure*}
All samples were fabricated in-house at the Nanofab facility of the Néel Institute. All lithography steps are done with an electron beam writer. We start from commercially available 500 nm thick LiNbO$_3$ (LN) on Si wafers and perform the fabrication process on smaller 12x12mm chips. We first thin down and pattern the LN using Ion beam etching (IBE) to a target thickness of $250$ nm. 
 We pattern the LN structures during a second IBE step, using a 1 µm-thick Medusa resist mask patterned by e-beam lithography.  After etching, the chips are cleaned using successive RCA and BOE processes to remove any LN and Medusa residues.  The wiring is then fabricated, starting with the Au ground planes and Al IDTs using standard lift-off techniques. The Al/AlOx/Al SQUIDs are fabricated using the bridge-free angled evaporation technique \cite{Planat2019}, and a second angled evaporation connects the various wiring sections while also climbing the LN mesa. Finally, the suspended LN structures are released using XeF$_2$ gas etching of the silicon substrate. To protect the chip during this process, resist is used to mask the device, and we patterned release windows only around the LN structures.
 This final step is performed after dicing the chips to their final $3 \times4$ mm size and just before measurement.
 After release, the samples are cleaned and bonded to a PCB and sample holder. Supplementary Fig.~\ref{figS3} shows the devices discussed in this paper at the end of the fabrication process.

\section{Measurement setup and calibration}

The samples are cooled down to cryogenic temperatures in a homemade dilution refrigerator. Scattering parameters, $S_{11}$ and $S_{21}$, are measured as a function of frequency using a network analyzer. A small coil generates magnetic fields of up to $\sim 3$ mT to thread flux in the SQUIDs. The microwave input is attenuated by a total of -83 dB, while the output is amplified by 75 dB over the $4$–$8$ GHz band. 
The microwave setup located on the mixing chamber at $\sim 30$ mK is shown in Supplementary Fig.~\ref{figS4}. We use a set of microwave switches inside the refrigerator to route the signal between multiple samples to be measured in reflection or transmission. A 6-way switch allows us to calibrate the reflection measurements using SOL standards. Standing waves of amplitude $\approx$ 1 dB remain from the coax cables running from the 6-way switch to the sample. Transmission measurements are calibrated by normalizing the data by the gain curve of the setup, which we obtain from measuring microwave throughs over several cooldowns with a systematic standard deviation of about 0.5 dB.

\section{Computation of the transducer scattering matrix}

An interdigital transducer can be modeled as a 2-port network with one microwave port (denoted by subscript $\mu$) and one acoustic port (denoted by subscript a). For a bi-directional IDT, we consider only the acoustic emitted in the forward direction, as the backward emitted waves will contribute only to losses. The scattering matrix of a transducer can then be written as:
\begin{equation}
    S_\text{IDT} = 
    \begin{pmatrix}
        S_{\mu \mu} &  S_{a \mu} \\
        S_{a \mu} &  S_{a a} 
    \end{pmatrix}
\end{equation}
The IDT is reciprocal but not necessarily lossless so the matrix is symmetric and non-unitary.

The IDTs are arranged in a delay line geometry. We therefore need to take into account the propagation of acoustics waves between the IDTs as well as the contribution of the multiple reflections that happen between them. The total measured transmission and reflection of the delay line can therefore be written as:
\begin{align}
    S_{21} &= S_{a \mu}^2 e^{ -(\alpha + i \beta) l} \left[1 +\sum_{n=1}^{+ \infty}\left( S_{a a}^{2} e^{ -(\alpha + i \beta) l}\right)^n \right]\\
    S_{11} &= S_{\mu \mu} + S_{a \mu}^2   \sum_{n=1}^{+ \infty}\left( S_{a a} e^{ -(\alpha + i \beta) 2 l}\right)^n 
\end{align}
where $\alpha$ and $\beta$ are the acoustic attenuation and propagation constants.
 When Fourier transforming the raw data to the time domain, we observe multiple peaks corresponding to these echoes, as shown in Supplementary Fig.~\ref{figS5}a. To analyze specific transits, we apply rectangular filtering windows to isolate these peaks and then Fourier transform the data back into the frequency domain. (Supplementary Fig.~\ref{figS5}b). 
 
 Thus we compute $S_{\mu \mu}$ from the first peak of $S_{11}$ and $S_{a \mu}$ from the first acoustic peak of $S_{21}$. $\lvert S_{a a} \rvert ^2$ can be computed from the transmission single and triple transits by noting that:
\begin{equation}
    \frac{\lvert S_{21}^{\text{3-T}}\rvert ^2}{\lvert S_{21}^{\text{1-T}}\rvert ^2} =   \frac{\lvert S_{a \mu}\rvert^2 \lvert S_{a a}\rvert^{2} e^{ -\alpha  3 l}}{\lvert S_{a \mu}\rvert^2 e^{ -\alpha l}} = \lvert S_{a a}\rvert^{2} e^{ -\alpha  l} 
\end{equation}
For our devices, acoustic propagation losses are small, with $e^{ -\alpha  l} = 0.97 $ for a typical 35 µm long delay line. Thus we approximate: $\lvert S_{a a} \rvert^{2} \approx \frac{\lvert S_{21}^{\text{3-T}}\rvert ^2}{\lvert S_{21}^{\text{1-T}}\rvert ^2}$.

The choice of filtering window introduces ambiguity, as it is not always clear how to define the start and end of a peak in the IDT's impulse response. This ambiguity is problematic, as variations in the width and position of the window can significantly alter the final values of the calculated scattering parameters. To quantify selection bias, we perform a statistical study by varying the position and size of the Fourier window for each peak. This process produces 900 samples for each scattering parameter of the IDT. For each parameter, we compute the mean value and 90\% confidence intervals. In the main text, we present this data in Fig.~2 and Fig.~3 for the magnetic flux at which the efficiency–bandwidth product is maximal.

\begin{figure}[t!]
    \includegraphics{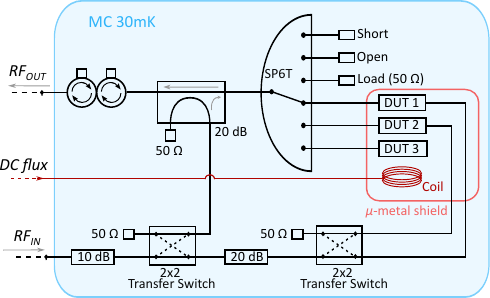}
    \caption{\label{figS4} \textbf{Cryogenic microwave measurement setup}}
\end{figure}

\begin{figure}[t!]
    \includegraphics{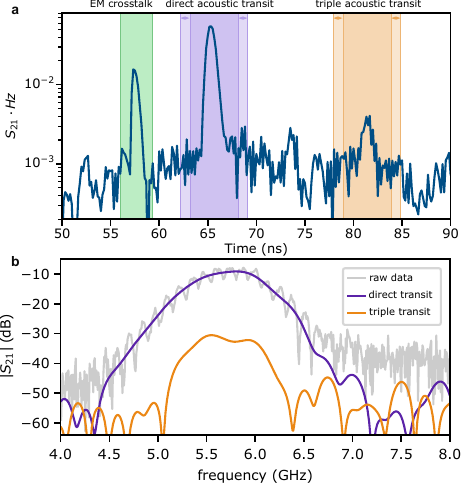}
    \caption{\label{figS5} \textbf{Fourier filtering of delay line transits}. \textbf{a.} Impulse response of the bi-directional IDT.  Shaded areas indicate the Fourier windows used to filter the single (purple) and triple transit (orange) contribution which are shown in panel \textbf{b.} after Fourier transforming back to frequency domain. }
\end{figure}
\begin{figure}[t!]
    \includegraphics{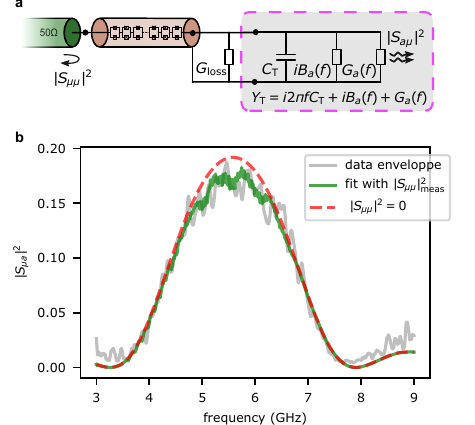}
    \caption{\label{figS6} \textbf{Fitting of $G_\text{loss}$.} \textbf{a.} Electrical model of the transducer including the parallel $G_\text{loss}$. \textbf{b.} Envelope of the transduction efficiency fitted taking measured reflection into account (green). The extracted $G_\text{loss}$ is then used to plot the red dashed curve supposing a perfect match.}
\end{figure}
\section{Circuit model for inefficiencies and tunability}
In this section we introduce the model used to explain the data of Fig.~4 of the main text.
The microwave power incident on the transducer can be reflected, transduced as acoustic waves, or dissipated in various loss channels. Thus, a perfectly matched IDT does not guarantee efficient transduction into the acoustic channel. We model the total contribution of losses as a conductance $G_\text{loss}$ in parallel with the IDT admittance, $Y_\text{IDT}$, as shown in  Supplementary Fig.~\ref{figS6} a. The efficiency of the IDT can be written as:
\begin{equation}
    \lvert S_{{a \mu}} \rvert ^2 = \left( 1 - \lvert S_{\mu \mu} \rvert^2 \right) \frac{G_\text{a}(\omega)}{G_\text{a}(\omega) + G_\text{loss}}
\end{equation}

We fit this expression to the envelope of the flux dependent $\lvert S_{a \mu} \rvert^2$ taking into account the measured $\lvert S_{ \mu \mu} \rvert^2$ (green curve of  Supplementary Fig.~\ref{figS6}). From this fit, we extract a value of $1/G_\text{loss} = 8.4 \text{ k} \Omega$ which we use to compute the transduction efficiency supposing perfect match at all frequencies (red dashed curve in Supplementary Fig.~\ref{figS6}, grey dashed fit curve in Fig.~4 of main text).
We can similarly fit $G_\text{loss}$ for all the different transducers (see Table~\ref{SMtab:loss}).
\begin{table*}[t!]
    \centering
    \setlength{\tabcolsep}{10pt} 
    \renewcommand{\arraystretch}{1.5} 
    \resizebox{\textwidth}{!}{ 
        \begin{tabular}{|c||c|c|c|c|c|c|c|c|c|}
        \hline
        \multirow{3}{*}{ } & \multicolumn{3}{c|}{Measured}   & \multicolumn{3}{c|}{Intrinsic losses} & \multicolumn{2}{c|}{Extrinsic losses} &  \multirow{3}{*}{Unexplained} \\
        \hhline{~--------~}
        \multirow{2}{*}{  } & Efficiency & Mismatch  & \multirow{2}{*}{ $1/G_\text{loss} $} & Backwards  & \multirow{2}{*}{SQUID losses} & \multirow{2}{*}{EM radiation} & Acoustic  & Beam &  \\
        &  $\lvert S_{a \mu} \rvert ^2$ & $\lvert S_{\mu \mu} \rvert ^2$ & & emitted power & & & propagation & steering&\\
        \hline\hline
        bi-directional IDT & 35 \% & 4 \% & $6 \text{ k} \Omega$ & 39.2 \% & 0.4 \%  &  $>6.5$ \% & 2.7 \%& 1.5 \% &  10.7 \%\\
        \hline
        UDT  & 61 \% & 16 \% & $7.4 \text{ k} \Omega$ & N/A & 0.7\% & $>3.8$ \% &  3.2 \% & 1.8 \% & 13.5\% \\
        \hline
        Single-mode & 25 \%  & 17.9 \% & $17.3 \text{ k} \Omega$ &  25 \% &  0.5 \% & $> 17.7$ \%  & Uncharaterized & N/A & 13.9 \%\\
        \hline
        Widely tunable IDT & 17.8 \% & 8 \% & $8.4 \text{ k} \Omega$ & 39.4 \% & 0.2 \% & $>3.2$ \% & 13.6 \% & 8 \%&   9.8 \%\\
        \hline
        \end{tabular}
    }
    \caption{\label{SMtab:loss} \textbf{Loss Budget.} Summary of all the different contributions to device inefficiencies.}
\end{table*}

\section{Loss budget}

We characterize some known loss mechanisms that contribute to the inefficiencies of our different transducers. 

First, we estimate the acoustic propagation losses that occur in the delay line between two transducers. Note that, these losses are not intrinsic to the transducer itself, and their contribution can therefore be excluded from $\lvert S_{a\mu} \rvert^2$ to determine the internal transducer efficiency.
We compute propagation losses by analyzing the resonances of the acoustic resonator shown in Supplementary Fig.~\ref{figS3}d). 
The attenuation constant, $\alpha$, is estimated from the internal loss rate, $\kappa_{\text{int}}$, using the relation:
\begin{equation}\label{SMeq:alpha}
    \alpha = \frac{\pi \kappa_{\text{int}}}{v} = \frac{\pi}{2 Q_\text{int} L}
\end{equation}
where $v$ is the phase velocity and $L$ the resonator length. 
Using equation \ref{SMeq:alpha} assumes that no losses occur at the mirrors, meaning that the $\alpha$ obtained from measurements includes the contribution of mirror-related losses. At an input power of $-108$ dBm, we find $\alpha = 915 \text{ m}^{-1}$ for plane waves. This value is then used to compute the power lost due to acoustic propagation for each device (see Supplementary Table~\ref{SMtab:loss}). 
In the case of single-mode devices, we were unable to make acoustic resonators with a clean enough response to characterize acoustic propagation losses. 
Since these propagation losses are derived from resonances, they do not include beam steering. The beam steering angle can be estimated from simulations to be 0.02 rad at the propagation angle shown in Supplementary Fig.~\ref{figS2}b. The corresponding losses are then computed and summarized in Supplementary Table~\ref{SMtab:loss}.

For bi-directional transducers, an equal amount of power is emitted in both directions.  To estimate the backward emitted power directly at the IDT output (before propagation starts), the extrinsic acoustic propagation losses are subtracted from the measured external efficiency $\lvert S_{a \mu} \rvert ^2$.

Similarly to acoustic losses, we estimate the losses that occur during the propagation of the microwave signal through the SQUID array. 
Independent measurements on $500 \text{ }\mu \text{m}$ long arrays give $Q_\text{int}=230 $. 
Using equation \ref{SMeq:alpha} this corresponds to an attenuation constant of $\alpha = 14 \text{ m}^{-1}$. For the typical $ \approx 100 - 500$ µm long arrays that we use as matching circuits, the resulting contribution to power loss is smaller than 1 \% (see Table~\ref{SMtab:loss}).

Finally, we estimate a lower bound on the power lost by electromagnetic radiation.
We compute the amount of power transmitted through the delay line as electromagnetic crosstalk by Fourier filtering the first peak of the delay line's impulse response (green window in Supplementary Fig.~\ref{figS5}). Assuming that all EM radiation emitted by one transducer is entirely picked up by the other as crosstalk, we obtain a lower bound on the energy lost as EM radiation.  

We summarize the computed values for all these various loss mechanisms in Table~\ref{SMtab:loss}. By accounting for all these contributions, we are able to explain the majority of the inefficiency observed in our devices.

\end{document}